\documentclass{emulateapj}

\begin{document}
\slugcomment{ApJ Letters, in press}

\shortauthors{Gordon et al.}
\shorttitle{SMC Tail Dust}

\title{The Dust-to-Gas Ratio in the Small Magellanic Cloud Tail} 

\author{K.~D.~Gordon\altaffilmark{1}, 
   C.~Bot\altaffilmark{2},
   E.~Muller\altaffilmark{3},
   K.~A.~Misselt\altaffilmark{4},
   A.~Bolatto\altaffilmark{5},
   J.-P.~Bernard\altaffilmark{6},
   W.~Reach\altaffilmark{7}
   C.~W.~Engelbracht\altaffilmark{4}, 
   B.~Babler\altaffilmark{8},
   S.~Bracker\altaffilmark{8},
   M.~Block\altaffilmark{4},
   G.~C.~Clayton\altaffilmark{9},
   J.~Hora\altaffilmark{10},
   R.~Indebetouw\altaffilmark{11},
   F.~P.~Israel\altaffilmark{12},
   A.~Li\altaffilmark{13},
   S.~Madden\altaffilmark{14},
   M.~Meade\altaffilmark{8},
   M.~Meixner\altaffilmark{1},
   M.~Sewilo\altaffilmark{1},
   B.~Shiao\altaffilmark{1},
   L.~J.~Smith\altaffilmark{1,16},
   J.~Th.~van~Loon\altaffilmark{15}, \&
   B.~A.~Whitney\altaffilmark{17}
   }

\altaffiltext{1}{STScI, 3700 San Martin Drive, Baltimore, MD 21218}
\altaffiltext{2}{UMR 7550, Centre de Donn\'ees Astronomique de Strasbourg (CDS), Universit\'e Louis Pasteur, 67000 Strasbourg, France}
\altaffiltext{3}{Dept.\ of Astrophysics, Nagoya Univ., Furo-cho, Chikusa-ku, Nagoya 464-8602, Japan}
\altaffiltext{4}{Steward Observatory, Univ.\ of Arizona, Tucson, AZ 85721}
\altaffiltext{5}{Dept.\ of Astronomy, Univ.\ of Maryland, College Park, MD 20742}
\altaffiltext{6}{CESR, 9 Av du Colonel Roche, 31028, Toulouse, France}
\altaffiltext{7}{IPAC, Caltech, MS 220-6, Pasadena, CA 91125}
\altaffiltext{8}{Dept.\ of Astronomy, Univ.\ of Wisconsin-Madison, 475 N. Charter St., Madison, WI 53706}
\altaffiltext{9}{Dept.\ of Physics \& Astronomy, Louisiana State Univ., Baton Rouge, LA 70803}
\altaffiltext{10}{Harvard-Smithsonian, CfA, 60 Garden St., MS 65, Cambridge, MA 02138}
\altaffiltext{11}{Dept.\ of Astronomy, Univ.\ of Virginia, P.O. Box 3818, Charlottesville, VA 22903}
\altaffiltext{12}{Sterrewacht Leiden, Leiden University, PO Box 9513, 2300 RA Leiden, The Netherlands}
\altaffiltext{13}{Dept.\ of Physics \& Astronomy, Univ.\ of Missouri, Columbia MO 65211}
\altaffiltext{14}{Service d'Astrophysique, CEA/Saclay, l'Orme des Merisiers, 91191 Gif-sur-Yvette, France}
\altaffiltext{15}{Astrophysics Group, Lennard-Jones Lab., Keele Univ., Staffordshire ST5 5BG, UK}
\altaffiltext{16}{Dept.\ of Physics \& Astronomy, Univ.\ College London, Grower St., London WC1E 6BT, UK}
\altaffiltext{17}{Space Sci.\ Inst., 4750 Walnut St., Suite 205, Boulder, CO 80301}

\begin{abstract} 
The Tail region of the Small Magellanic Cloud (SMC) was imaged using
the MIPS instrument on the Spitzer Space Telescope as part of the
SAGE-SMC Spitzer Legacy.  Diffuse infrared emission from dust was
detected in all the MIPS bands.  The Tail gas-to-dust ratio was
measured to be $1200 \pm 350$ using the MIPS observations combined
with existing IRAS and HI observations.  This gas-to-dust ratio is
higher than the expected 500--800 from the known Tail metallicity
indicating possible destruction of dust grains.  Two cluster regions
in the Tail were resolved into multiple sources in the MIPS
observations and local gas-to-dust ratios were measured to be
$\sim$440 and $\sim$250 suggests dust formation and/or significant
amounts of ionized gas in these regions.  These results support the
interpretation that the SMC Tail is a tidal tail recently stripped
from the SMC that includes gas, dust, and young stars.
\end{abstract}


\section{Introduction}
\label{sec_intro}

Among the nearby galaxies, the Small Magellanic Cloud (SMC) and nearby
regions of the Magellanic Bridge represent a unique astrophysical
laboratory for interstellar medium (ISM) studies, because of the SMC's
proximity \citep[$\sim$60 kpc,][]{Hilditch05}, low metallicity
\citep[1/5-1/8 Z$_{\sun}$,][]{Russell92, Rolleston99, Lee05,
Perez-Montero05} and tidally-disrupted interaction status
\citep{Zaritsky04}.  The SMC offers a rare glimpse into the physical
processes in an environment with a metallicity that is below the
threshold of 1/3--1/4 Z$_{\sun}$ where the properties of the ISM change
as traced by the rapid reduction in the PAH dust mass fractions and
dust-to-gas ratios \citep{Engelbracht05SB, Draine07}.  In addition,
the SMC is the only local galaxy that has the ultraviolet dust
characteristics \citep[lack of 2175~\AA\ extinction bump,][]{Gordon03}
of starburst galaxies in the local \citep{Calzetti94, Gordon97} and
high-redshift \citep[$2<z<4$,][]{Vijh03} universe.  The Large and
Small Magellanic clouds represent the nearest examples of tidally
interacting galaxies and the Magellanic Bridge is a clear
manifestation of a close encounter of these two galaxies some 200 Myr
ago \citep{Zaritsky04}.

\begin{figure*}[tbp]
\epsscale{1.1}
\plotone{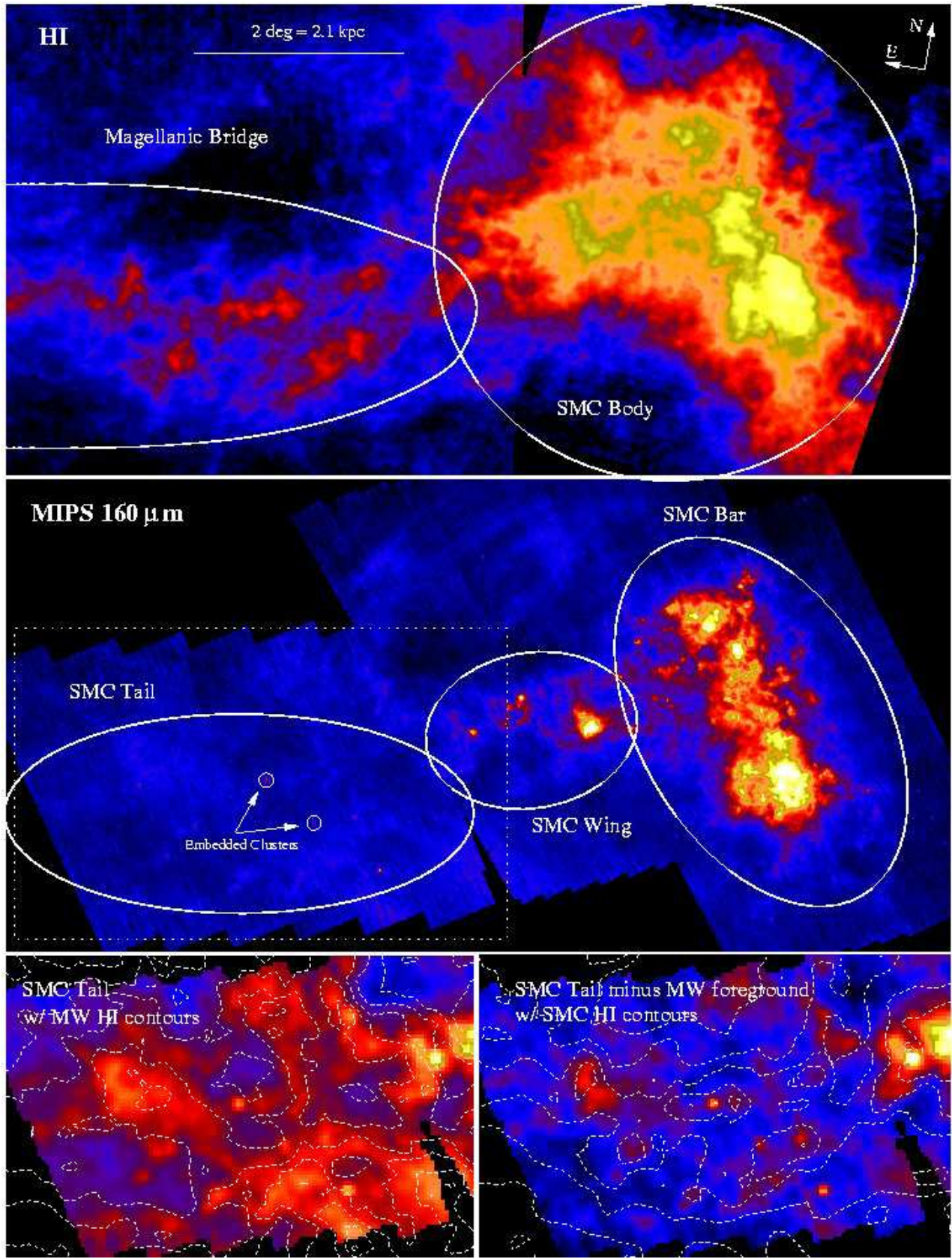}
\caption{The full SMC in HI (top panel) and at 160~\micron\ (middle
panel) is shown and various regions are labeled.  The dotted box in
the middle panel indicates the region shown in the two bottom panels.
The left bottom panel shows the 160~\micron\ data convolved to the
IRIS-reprocessed IRAS 100~\micron\ resolution ($4\farcm 3$) overlaid
with MW HI contours.  The right bottom panel displays the same
region after subtraction for the predicted MW foreground cirrus
emission.  The contours on this panel show the SMC HI measurements.
Both bottom panels are displayed on the same scale.
\label{fig_smc_full} }
\end{figure*}

In particular, the SMC Tail (Fig.~\ref{fig_smc_full}) represents one
of the nearest example of tidally stripped material.  We define the
SMC Tail as the portion of the Magellanic Bridge that is adjacent to
the SMC Wing and has a higher density and metallicity than the rest of
Magellanic Bridge.  The Magellanic Bridge (which includes the SMC
Tail) is a filament of neutral hydrogen, which joins the SMC and LMC
over some 15~kpc \citep{McGee86, Stavely-Smith98}.  \citet{Harris07}
found only locally formed, young ($<$200 Myrs) massive stars
associated with the SMC Tail.  There is a transition in the
metallicities between the SMC Tail with 1/5--1/8 Z$_{\sun}$ and nearby
Magellanic Bridge regions (east of the SMC Tail) with 1/20 Z$_{\sun}$
\citep{Rolleston99, Rolleston03, Lee05}.

The nature of the SMC Tail as a tidally stripped region with only
recent star formation makes the detection and measurement of the
amount of dust in this region important.  Is there dust in this region
of low HI
column density and metallicity?  Is the gas-to-dust ratio
the same as the Body/Wing or has the dust been destroyed due the harsh
environment in the Tail?  Is the gas-to-dust ratio consistent with
expectations for the low metallicity \citep{Draine07}?  The presence
of dust in the SMC Tail region has been inferred for select regions
from IRAS point sources.  The detection and measurement of widespread
dust in the SMC Tail requires sensitive far-infrared ($>$100~\micron)
observations where the bulk of the dust emission occurs.

The SAGE-SMC (Surveying the Agents of Galaxy Evolution in the
Tidally-Disrupted, Low-Metallicity Small Magellanic Cloud) Legacy
program is using the Spitzer Space Telescope \citep{Werner04Spitzer}
to map the SMC from 3.6--160~\micron\ with the Infrared Array Camera
\citep[IRAC,][]{Fazio04IRAC} and the Multiband Imaging Photometer for
Spitzer \citep[MIPS,][]{Rieke04MIPS}.  The SAGE-SMC observations cover
$\sim$$30$~$\sq\arcdeg$ and builds on the existing S$^3$MC
\citep{Bolatto07} observations of the central $\sim$$3$~$\sq\arcdeg$.
The SAGE-SMC observations consist of two complete maps taken in two
epochs.  The full details of the entire SAGE-SMC program are given by
Gordon et al. (in prep.).  This Letter presents the analysis of the
SAGE-SMC epoch 1 MIPS observations of the SMC Tail region.

\section{Data}
\label{sec_data}

The MIPS observations at 24, 70, and 160~\micron\ were taken on
15-23/9/2007 split into 22 separate maps.  Each map consisted of fast
rate scan legs with a cross scan offset of 148\arcsec\ and lengths
optimized for each map.  We used the MIPS Data Analysis Tool v3.06
\citep{Gordon05DAT} to do the basic processing and final mosaicking of
the individual images.  Extra processing steps were carried out
similar to those for SAGE-LMC \citep{Meixner06}.  The epoch 1 SAGE-SMC
MIPS 160~\micron\ observations are shown in Fig.~\ref{fig_smc_full}
along with the integrated HI column density for the same region
\citep{Stanimirovic99, Muller03}.  The MIPS 70 and 160~\micron\
mosaics were supplemented in SMC Body region with the S$^3$MC data
\citep{Bolatto07}.  The full details of the SAGE-SMC processing are
given by Gordon et al. (in prep.).

We also used the IRIS reprocessing of the 60 and 100~\micron\ IRAS
images \citep{Miville-Deschenes05} of the same region for our
analysis.  We convolved the MIPS images to the same resolution as the
IRIS-reprocessed IRAS 100~\micron\ image (FWHM = $4\farcm 3$) using
kernels created in the manner described in \citep{Gordon08}.

Emission from Milky Way (MW) foreground cirrus clouds dominates the
emission seen in the MIPS 160~\micron\ band in the SMC Tail
region. This is clearly seen in Fig.~\ref{fig_smc_full} in the lower
left panel where the 160~\micron\ image is shown with contours from
the integrated MW velocity HI gas \citep{Muller03} overplotted.  The
MIPS 160~\micron\ and MW HI gas emission are well correlated.  As was
done for the LMC \citep{Bernard08}, the MW HI gas emission can be used
to quantitatively predict the MW dust emission that can then be
subtracted from the observed MIPS 160~\micron\ (and other IR bands).
This reveals the SMC dust emission and is illustrated in the lower
right panel of Fig.~\ref{fig_smc_full} showing the MW foreground
subtracted MIPS 160~\micron\ image of the SMC Tail region overlaid
with the contours from the integrated SMC velocity HI gas
\citep{Muller03} overplotted.

\section{Results}

\begin{figure*}[tbp]
\epsscale{1.1}
\plotone{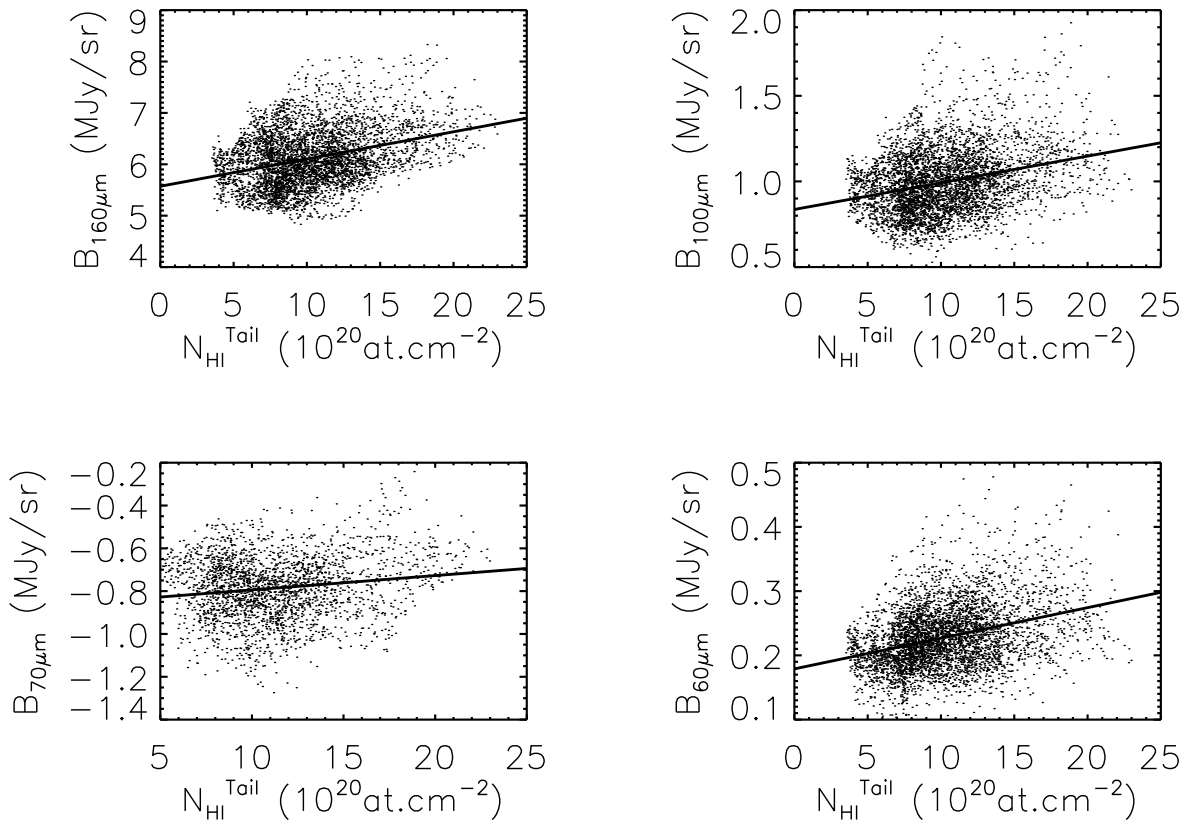}
\caption{The correlation between HI column in the Tail region,
$N_{HI}^{Tail}$, and the MW foreground subtracted MIPS/IRAS surface
brightness is shown.  The solid line gives the best fit linear
correlation.  The region over which the correlation is done was
restricted to that labeled SMC Tail in Fig.~\ref{fig_smc_full} except
at MIPS 70~\micron.  The MIPS 70~\micron\ correlation was done in a
smaller region centered on the full region to avoid having the
instrumental residuals bias the resulting correlation.  The negative
70~\micron\ surface brightnesses are artifacts of the processing for
this band where a smoothly varying background (in time) is subtracted
for each 70~\micron\ pixel to remove variations in the detector
baseline.  The significantly larger scatter of the MIPS 70~\micron\
points, as compared to the IRAS 60~\micron\ points, is an indication
there still are significant residual MIPS 70~\micron\ instrumental
signatures at these low surface brightness levels. \label{fig_correl}
}
\end{figure*}

The presence of dust in the SMC Tail region can be inferred from
Fig.~\ref{fig_smc_full} (bottom right panel).  Quantifying the amount
of dust can be done by correlating the MW foreground corrected SMC IR
emission with the SMC HI column densities.  Such correlations are
shown in Fig.~\ref{fig_correl} for the MIPS 70 and 160~\micron\ and
IRAS 60 and 100~\micron\ bands.  The MW foreground IR emission was
subtracted from the images by scaling the MW HI image using the
coefficients given in Table~\ref{tab_correl}.  There is a clear
correlation between these IR emissions and the SMC HI column densities
at each wavelength and the slopes of these correlations are given in
Table~\ref{tab_correl}.  The point sources in the IR images were
removed using standard sigma clipping to avoid biasing the measurement
of the IR to HI correlation.  The slopes and formal uncertainties were
computed using two different methods that produced consistent
results.  To account for uncertainties in the foreground IR emission
subtraction and absolute calibrations, we have conservatively set the
slope uncertainties to 20\% for all but the MIPS 70~\micron\ band
where the uncertainty is 30\% to account for the use of the smaller
region.  These uncertainties are 
well above the formal uncertainties in the correlation fits.

\begin{deluxetable}{rcc}
\tablewidth{0pt}
\tablecaption{Infrared to HI Slopes \label{tab_correl}}
\tablehead{  & \colhead{MW\tablenotemark{a}} & \colhead{SMC Tail} \\
  \colhead{Band} & 
  \multicolumn{2}{c}{[MJy sr$^{-1}$ (10$^{20}$ atoms cm$^{-2}$)$^{-1}$]} }
\startdata
IRAS 60  & 0.129 & $0.0048 \pm 0.0010$ \\
MIPS 70  & 0.191 & $0.0067 \pm 0.0020$ \\
IRAS 100 & 0.522 & $0.016 \pm 0.0032$ \\
MIPS 160 & 0.971 & $0.053 \pm 0.011$ \\
\enddata
\tablenotetext{a}{Adopted from \citet{Bernard08}.}
\end{deluxetable}

The SMC Tail HI gas-to-dust ratio can be computed from the IR to HI
slopes assuming a dust grain model.  The 60 and 70~\micron\ emission
can include contributions from non-equilibrium emission processes
\citep{Bernard08} and so we only use the 100 and 160~\micron\
correlations to determine the dust temperature.  We compute a dust
temperature of $15 \pm 1.1$~K using a black body modified by the
emissivity of silicate grains \citep[$\propto
\lambda^{-2}$,][]{Weingartner01}.  Including the 60 and 70~\micron\ 
correlations increases the temperature to 21~K clearly indicating
significant non-equilibrium emission at these wavelengths.  The
computed HI gas-to-dust ratio is $1200 \pm 350$ using T~=~15~K.  For
reference, this method gives a HI gas-to-dust ratio of 84 for the MW
slopes given by \cite{Boulanger96} (interpolating to get the
160~\micron\ slope).  This MW gas-to-dust ratio is similar to the
standard 100-110 value
\citep{Sofia01, Sofia01errat, Bot07, Draine07}.  We adopted a MW
reference value of 100, the SMC Tail HI
gas-to-dust ratio is ($12 \pm 3.5$)$\times$ higher than the MW ratio.
An analysis using the method of \citet{Bot04} gives a HI gas-to-dust
ratio that is 12$\times$ MW.  This factor is higher than the expected
gas-to-dust ratio value of 5-8$\times$ MW for a metallicity that is
1/5-1/8 solar \citep{Draine07}.  It is very unlikely that this
discrepancy between the measured and expected ratio is due to not
accounting for the molecular gas component.  The gas in the SMC Tail
region has a measured molecular-to-atomic ratio of 0.002
\citep{Mizuno06}. 

\subsection{Embedded Clusters}

\begin{figure*}[tbp]
\epsscale{1.1}
\plotone{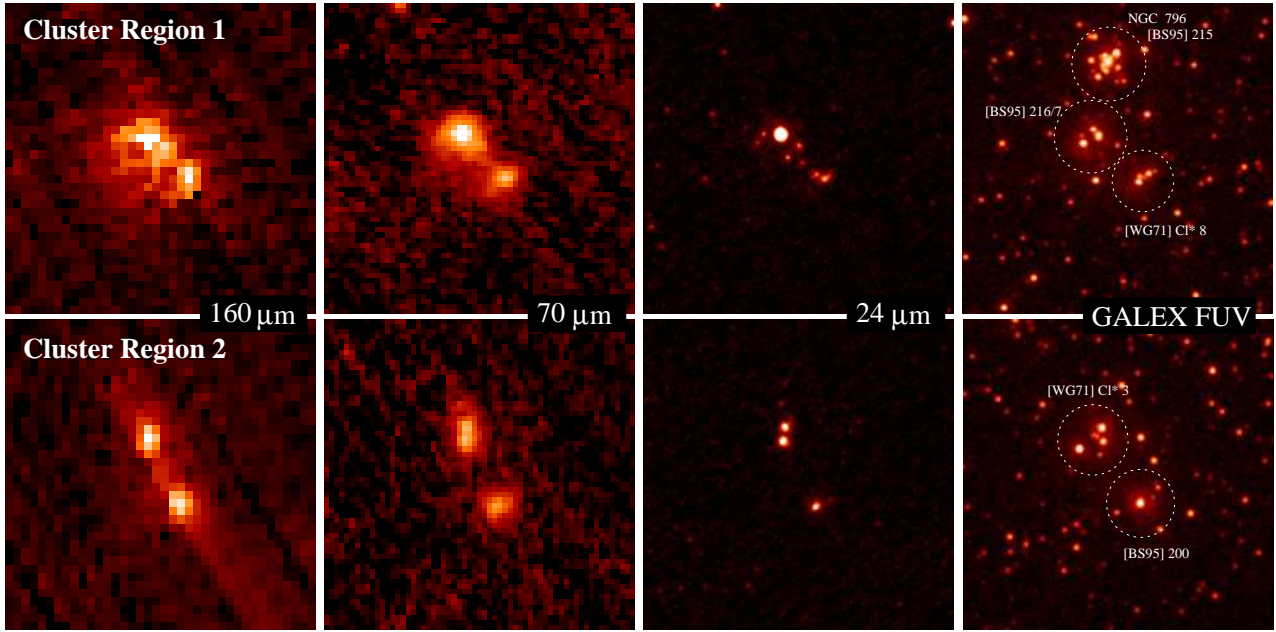}
\caption{The IR appearance of the two embedded clusters is shown for
$10\arcmin \times 10\arcmin$ regions.  These two regions correspond to
regions C and B of \citet{Mizuno06}.  In addition, the GALEX far-UV
images of the same regions are shown illustrating the young nature of
these objects.  The streaking seen in the 160~\micron\ and (to a
lesser extent) 70~\micron\ images are residual instrument artifacts.
\label{fig_clusters_cutouts} }
\end{figure*}

There are two regions of enhanced, extended emission at 160~\micron\
in the SMC Tail region.  These two regions are resolved into multiple
sources in all the MIPS bands.  The locations of these embedded
clusters are shown in Fig.~\ref{fig_smc_full} and closeups of the
clusters are shown in Fig.~\ref{fig_clusters_cutouts}.  The locations
of known clusters \citep{Westerlund71, Bica95} in these regions are
shown on archival GALEX far-ultraviolet images.  The simultaneous
presence of far-UV and far-IR emission indicates young, embedded star
forming regions.  These cluster regions also have strong H$\alpha$
emission \citep{Muller07} providing the indication that they harbor
massive star formation.  The IR stellar properties of these clusters
will be investigated in more detail in Chen et al.\ (in prep.).

\begin{deluxetable}{lcc}
\tablewidth{0pt}
\tablecaption{Cluster Properties \label{tab_clust_prop}}
\tablehead{\colhead{Property} & \colhead{Region 1} & \colhead{Region 2}}
\startdata
RA(2000) & 01 56 49 & 01 49 34.39 \\
DEC(2000) & -74 16 24 & -74 38 12 \\
MIPS 24 [Jy] & $0.153 \pm 0.017$ & $0.062 \pm 0.005$ \\
IRAS 25 [Jy] & $0.215 \pm 0.005$ & $0.187 \pm 0.007$ \\
IRAS 60 [Jy] & $3.47 \pm 0.05$ & $1.19 \pm 0.02$ \\
MIPS 70 [Jy] & $4.94 \pm 0.02$ & $2.15 \pm 0.01$ \\
IRAS 100 [Jy] & $5.80 \pm 0.03$ & $2.81 \pm 0.008$ \\
MIPS 160 [Jy] & $9.60 \pm 0.01$ & $5.08 \pm 0.005$ \\
Dust Temp. [K] & $16.5^{+3.5}_{-2.0}$ & $17.5^{+4.4}_{-2.5}$ \\
Dust Mass [M$_{\sun}$] & $320^{+700}_{-150}$ & $120^{+290}_{-90}$ \\
HI Mass [M$_{\sun}$] & $8.0 \times 10^{4}$ & $2.2 \times 10^{4}$ \\
HI Gas/Dust & $250^{+220}_{-170}$ & $183^{+553}_{-126}$ \\
H2 Mass [M$_{\sun}$] & $7 \times 10^{3}$ & $1 \times 10^{3}$ \\
(HI+H2) Gas/Dust & $\sim 440$ & $\sim 250$ \\
\enddata
\end{deluxetable}

These embedded clusters are a clear indication of very recent star
formation and provide probes of the local gas-to-dust ratio.  The dust
mass in each cluster was determined from IR fluxes given in
Table~\ref{tab_clust_prop}.  The IR fluxes were measured using using 
an aperture with a radius of $6\farcm 67 \sim 116$~pc and a sky
annulus with min/max radii of $8\farcm 33$/$11\farcm 67$.  The
uncertainties were calculated using the measured noise in the sky
annulus.  The dust masses were determined by fitting a two component
(warm and cold silicates) grain model to the IR fluxes
\citep{Marleau06}.  The cold silicate (which dominates the dust mass)
dust temperatures and corresponding dust masses are given in
Table~\ref{tab_clust_prop} along with the HI masses determined using
the same aperture.  Employing the the same analysis described above
for the SMC Tail yields dust masses similar to those in
Table~\ref{tab_clust_prop}.  The HI gas-to-dust ratios are smaller
than for the whole SMC Tail, suggesting local dust enhancements.  One
alternate explanation for the lower gas-to-dust ratio would be a
variable molecular gas content in the SMC Tail region.  Adding the
H$_2$ contribution from the measured CO fluxes
\citep{Muller03CO, Mizuno06}, adjusted for a more appropriate
CO-to-H$_2$ conversion ratio \cite{Israel97}, gives total gas-to-dust
ratios (Table~\ref{tab_clust_prop}) that are still below that measured
for the whole SMC Tail.

\section{Summary/Discussion}

We have detected the diffuse infrared emission from dust in the SMC
Tail portion of the Magellanic Bridge using the epoch 1 MIPS SAGE-SMC
Spitzer Legacy observations.  The gas-to-dust ratio in the SMC Tail
region was measured to be $1200 \pm 350$, ($12 \pm 3.5$)$\times$ MW
value, using correlations between the infrared emission and HI column
densities.  This value is in reasonable agreement with the range of
measured SMC Body gas-to-dust ratios
\citep[5-11$\times$ MW,][]{Gordon03, Bot07, Leroy07SMC} determined from UV
extinction and IR emission measurements.  This is consistent with the
picture that the SMC Tail has been recently stripped from the SMC Body
during a tidal encounter \citep{Connors06} and evidence that the gas
and dust in the SMC Tail have not been stripped from the LMC or are
due to infalling material from the inter-galactic medium.  Looking more closely, our
infrared measured SMC Tail gas-to-dust ratio is higher than the SMC
Body gas-to-dust ratio of 7$\times$ MW \citep{Leroy07SMC} that was
determined using IR emission measurements and a similar method.  In
addition, the SMC gas-to-dust ratio of 12$\times$ MW is higher than
the expected value from of 5-8$\times$ MW \citep{Draine07} for the
1/5-1/8 solar metallicity of the SMC Tail region \citep{Lee05}.  This
may indicate there has been destruction of the dust in the SMC Tail or
that there is colder dust that is not detected by the MIPS
160~\micron\ observations.  Possible plausible dust destruction
mechanisms include a harder radiation field due to less dust
shielding and shocks due to the tidal interaction \citep{Jones96}.

Two cluster regions are detected in the MIPS observations indicating
the presence of young, embedded star formation.  The local gas-to-dust
ratio was measured for these two regions to be 2.5--4.4$\times$ MW.
These ratios are lower than the gas-to-dust ratio measured for the
entire SMC Tail (12$\times$ MW) and lower than that expected for the
metallicity of the SMC Tail (5-8$\times$ MW).  This suggests there has been
dust formation and/or a significant amount of ionized gas is present
in these regions.  As both regions harbor known HII regions
\citep{Muller07}, the latter is clearly part of the answer.

This letter presents the detection of dust in the SMC Tail and a
preliminary analysis of its dust-to-gas ratio.  A more complete
analysis of the dust in the whole SMC (Bar, Wing, and Tail) will be
presented by Bot et al. (in prep.).

\acknowledgements
This work is based on observations made
with the {\em Spitzer Space Telescope}, which is operated by the Jet
Propulsion Laboratory, California Institute of Technology under NASA
contract 1407. Support for this work was provided by NASA through
a contract issued by JPL/Caltech to Space Telescope Science Institute.


\end{document}